\documentstyle[12pt]{article}
\begin{document}

\title{Fermion Quasi-Spherical Harmonics}

\author{\thanks{Author to whom correspondence should be addressed;
email: ghunter@yorku.ca}
G.Hunter, P.Ecimovic, I.Schlifer, I.M.Walker,
D.Beamish, \\
\thanks{Permanent Address: Institute for
Nuclear Energy Research, Bulgarian Academy of Sciences, Sofia, Bulgaria}
S.Donev, M.Kowalski, 
S.Arslan and S.Heck.\\  
{\small\sl Centre for Research in Earth and Space Science},\\
{\small\sl York University, Toronto, Canada  M3J 1P3}}


\maketitle

\vspace*{-0.25in}

\begin{abstract}
Quasi-Spherical harmonics, 
$Y_\ell^{m}(\theta,\phi)$ 
are derived and presented for
half-odd-integer values of $\ell$ and $m$.  The form of the $\phi$ factor 
is identical to that in the case of integer $\ell$ and $m$: 
$\exp{({\rm i} m \phi)}$.  However, the domain of these functions in the
half-odd-integer case is $0$$\leq$$\phi$$<$$4\pi$ rather than the domain
$0$$\leq$$\phi$$<$$2\pi$ in the case of integer 
$\ell$ and $m$ (the true spherical harmonics).
The form of the $\theta$ factor, $P^{|m|}_\ell(\theta)$ 
(an associated Legendre function) is  
(as in the integer case) the factor $(\sin\theta)^{|m|}$ multiplied by a
polynomial in $\cos\theta$ of degree $(\ell\rm{-}|m|)$ (an associated 
Legendre polynomial).  
A substantial difference
between the spherical (integer $\ell$ and $m$) and quasi-spherical 
(half-odd-integer $\ell$ and $m$) Legendre functions is that 
the latter have an irrational factor of $\sqrt{\sin\theta}$ whereas 
the factor of the truly spherical functions is 
an integer power of $\sin\theta$.
The domain of both the true and quasi spherical
associated Legendre functions is the same: $0$$\leq$$\theta$$<$$\pi$.  
A table of the Associated Legendre Functions is presented for both 
integer and 
half-odd-integer values of $\ell$ and $m$, for $|m|= 0, \frac{1}{2}, 
1 \ldots \frac{11}{2}$, 
and for $(\ell\rm{-}|m|)=0, 1, 2, 3, 4, 5$.  
The table displays the similarity
between the functions for integer $\ell$ and $m$ (which are well known) 
and those for half-odd-integer $\ell$ and $m$ 
(which have not been recognized previously).
\end{abstract}


\section{Introduction}

The theory of angular momentum 
based upon the fundamental commutation relations 
\cite[p.93]{Rae} 
\cite[p.309]{Kaempffer}, \cite[pp.107-112]{Hannabuss}, 
produces the 
eigenfunctions of 
$\:{\bf L^2}\:$ (the square of the total angular momentum) 
and $\:{\bf L_z}\:$ (its $z$-component) with eigenvalues 
of $\:\hbar^2 \ell(\ell+1)\:$
and $\:\hbar m$, respectively 
\cite[\S 5.4 pp.100-103]{Rae}.  
This general theory 
 leads (via the
raising and lowering ladder operators) to the prediction of both integer and
half-odd-integer values of the quantum numbers $\:\ell\:$ and $\:m\:$
\cite[\S 5.4 pp.100-103]{Rae}, \cite[p.311]{Kaempffer}, with
the manifold of eigenvalues defined by:
\begin{itemize}
\item
$\ell= 0\;\: \frac{1}{2}\;\: 1\;\: \frac{3}{2}\;\: 
2\;\: \frac{5}{2}\;\: 3\;\: \ldots$
\item $m=-$$\ell\;\: $$-$$(\ell$$-$$1)\;\: $$-$$(\ell$$-$$2)$$\;\: \ldots $
$+$$(\ell$$-$$2)$$\;\: +$$(\ell$$-$$1)$$\;\: +$$\ell\;$ for given $\:\ell\:$.
\end{itemize}
The quantum numbers, $\:\ell\:$ and $\:m\:$, are half-odd-integers
for 
the spin and total (spin+orbital) angular momentum of fermions, and are  
integers 
for fermion orbital angular momentum, and for boson spin and
total angular momentum.  

This general theory of angular momentum is {\em abstract}
in the sense 
that the operators, 
$\:{\bf L^2}\:$ and $\:{\bf L_z}\:$, and their 
eigenfunctions, are not functions of any coordinates.

In the Schr\"odinger wave mechanics of the hydrogen atom
the operators, $\:{\bf L^2}$ and $\:{\bf L_z}$ (that represent the
motion of the electron around the proton) are expressed 
 in terms of spherical polar coordinates: 
$\:r, \:\theta, \:\phi\:$ \cite[p.207]{McQuarrie},\\
\cite[p.95]{Rae}, where they have the form:
\begin{eqnarray}\label{amops} 
{\bf L^2} = - \hbar^2
\left\{
\frac{1}{\sin \theta }\frac{\partial}{\partial\theta}\left(\sin \theta 
\frac{\partial}{\partial\theta}\right) +
\frac{1}{\sin^2 \theta }\frac{\partial^2}{\partial\phi^2}
\right\} \quad\quad
{\bf L_z} = \frac{\hbar}{\rm i}\frac{\partial}{\partial\phi} 
\end{eqnarray}
The eigenfunctions in this coordinate representation are called 
spherical harmonics and denoted by  
$\:Y_\ell^m(\theta,\phi)$.
This Schr\"odinger representation of angular momentum leads to the same
manifold of eigenvalues as the general theory, except that 
$\:\ell\:$ and $\:m\:$ are restricted to be integers 
\cite[pp.313-315]{Kaempffer}.  This restriction arises from
the argument that $\:Y_\ell^m(\theta,\phi)$
should be a single-valued function of the coordinates
\cite[p.103]{Rae}.

The purpose of this paper is to derive and present the eigenfunctions of 
the coordinate-representation operators (\ref{amops})
corresponding to half-odd-integer values of 
$\:\ell\:$ and $\:m$.  
Their $\:\theta\:$ factors are compared with those of the
 well-known eigenfunctions with integer 
$\:\ell\:$ and $\:m\:$ in Table \ref{legyfunc} on page \pageref{legyfunc}.

Notwithstanding their algebraic similarity 
(apparent from Table \ref{legyfunc}), the integer and half-integer 
functions have different domains: the integer functions are defined on
a Euclidean sphere, whereas the half-integer functions are not.  One way of
of interpreting this difference is that the angle $\:\phi\:$ has a range of
$\:0$$\leq$$\phi$$<$$4\pi\:$ for the half-integer functions compared with
$\:0$$\leq$$\phi$$<$$2\pi\:$ for the integer functions.
  This difference in domain is related to the 
essential difference
between orbital angular momentum and spin angular momentum.
The interpretation of the half-odd-integer functions is discussed in
 \S\ref{interp} on page \pageref{interp}.

Archival presentations of 
the associated 
Legendre functions \cite[p.332]{AandS} 
involve generating functions and general formul\ae\ which in 
principle allow explicit expressions and numerical values to be 
obtained for any required values of $\:\ell\:$ and $\:m\:$. 
However some of these formul\ae\ become undefined
when $\:\ell\:$ and $\:m\:$ are not integers, because they involve 
differentiation (with respect to $\:\theta$) 
$\:\ell\:$ or $\:m\:$ times; e.g.~Rodrigues' Formula 
\cite[\S 8.6.18 p.334]{AandS}.  

Thus to facilitate a clear exposition, we derive
 these functions as solutions of the appropriate differential equations.

\section{The $\:\theta\:$ and $\:\phi\:$ Differential Equation}

The differential equation  for the eigenfunctions, $\:Y(\theta,\phi)$
and eigenvalues, A, 
of the square of the total angular momentum, $\:{\bf L^2}\:$ is:
\begin{eqnarray}\label{TotalAM} 
- \left\{
\frac{1}{\sin \theta }\frac{\partial}{\partial\theta}\left(\sin \theta 
\frac{\partial}{\partial\theta}\right) +
\frac{1}{\sin^2 \theta }\frac{\partial^2}{\partial\phi^2}
\right\} Y(\theta,\phi) = A Y(\theta,\phi)
\end{eqnarray}
It is well-known that the independent variables, $\:\theta\:$ 
and $\:\phi\:$,
are separable, and hence $\:Y(\theta,\phi)\:$ can be written 
as a product:
\begin{equation}\label{TFP}
Y(\theta,\phi) = \Theta(\theta)\times \Phi(\phi)
\end{equation}
This leads to the separated, ordinary differential equations: 
\begin{eqnarray}\label{HLATeqn}
\left\{
{\sin \theta }\frac{\rm d}{{\rm d}\theta}\left(\sin \theta 
\frac{\rm d}{{\rm d}\theta}\right) + A\, \sin^2\theta - B
\right\} \Theta(\theta) = 0
\end{eqnarray}
and
\begin{eqnarray}\label{HLAFeqn}
\left\{\frac{{\rm d}^2}{{\rm d}\phi^2} + B
\right\} \Phi(\phi) = 0
\end{eqnarray}
where $\:B\:$ is the separation constant arising from the separation 
of $\:\theta$
from $\:\phi$.  

The solutions of (\ref{HLAFeqn}) are obvious (and well-known) to 
 have the form:
\begin{eqnarray}\label{Fsoln}
\Phi(\phi) = \exp({\rm i} m \phi) 
\end{eqnarray}
in which $\: m \:$ is a constant to be determined.
Substitution of (\ref{Fsoln}) into (\ref{HLAFeqn}) yields:
\begin{equation}\label{F1}
\left(-  m ^2 + B\right) \Phi(\phi) = 0 \quad\quad
{\rm or\ since\ }\Phi(\phi) \neq 0 \quad\quad
 m^2 = B
\end{equation}
This relationship between the exponent $\:m\:$ and the separation 
constant $\:B$ is necessary, but in itself it does not specify 
the solutions any further.
It is noteworthy that $\:B\:$ has {\em the same value} for the 
two different
solutions of (\ref{HLAFeqn}) having $\:m\:$ values equal in magnitude 
but
opposite in sign; e.g.~for $\:m$=+$2\:$ and $\:m$=$-$$2$, $\:B$=$4$.  

In the case of orbital angular momentum
 one proceeds by noting that the angle 
$\phi\:$  takes any value within a
complete circle 
($0\leq\phi\leq 2\pi$), 
and hence the appropriate condition is that the function 
$\:\Phi(\phi)\:$ shall have the same value when 
$\phi\:$ transits a complete circle \cite[pp.208-209]{McQuarrie},
\cite[p.38]{Hannabuss}:
\begin{equation}\label{F2}
\Phi(\phi + 2\pi) = \Phi(\phi) \quad
\exp({\rm i} m [\phi+2\pi]) = \exp({\rm i} m \phi) \Rightarrow 
\exp({\rm i} m  2\pi) = 1
\end{equation}
This requirement is necessary for $\:\Phi(\phi)\:$ to 
be a proper (i.e.~single-valued) function of the points that 
comprise the
surface of a sphere; it is met as long as $\: m \:$ is any integer.

We relax this single-valuedness condition on 
$\Phi(\phi)\:$ 
by leaving $\:m\:$ undefined at this stage of the
analysis, except that $\:m\:$ should be real in order for 
$\Phi(\phi)\:$ to be periodic (and hence non-singular).
Relaxing the traditional
single-valuedness condition on
$\Phi(\phi)\:$, leads, however, to a different, non-classical 
interpretation of the angle $\phi$ (see \S \ref{interp} on page 
\pageref{interp}).

\subsection{Solution of the $\:\theta\:$ Equation}

Having solved the $\:\phi\:$ differential equation and determined 
that the
separation constant $\:B\:$  has the value
$B$=$m^2$, the $\:\theta\:$ differential equation 
(\ref{HLATeqn}) (after dividing by $\:\sin^2\theta$) becomes:
\begin{equation}\label{SEtheta}
\frac{1}{\sin\theta}\frac{\rm d}{{\rm d}\theta}\left(\sin\theta 
\frac{{\rm d}\Theta}{{\rm d}\theta}\right) + 
\left[A - \frac{m^2}{\sin^2\theta}\right]\Theta
 = 0
\end{equation}
Following the usual derivation, we transform (\ref{SEtheta}) to
a new independent \\ variable $\:x\:$ defined by:
\begin{equation}\label{thetatox}
x = \cos\theta\quad\quad\Rightarrow\quad\quad\sin\theta = \sqrt{1-x^2}
\end{equation}
and rename the dependent variable  $\:P(x)$; 
i.e.~$\Theta(\theta)\equiv P(x)$.  The range of
$\theta$, $\:0 \leq\theta\leq \pi\:$ becomes $\:+1 \geq x\geq -1$.

With this transformation the differential equation (\ref{SEtheta}) 
becomes:
\begin{equation}\label{SEx}
(1-x^2)\frac{{\rm d}^2P}{{\rm d}x^2} - 2 x \frac{{\rm d}P}{{\rm d}x}
 + \left[A -  
\frac{m^2}{1-x^2}\right]P = 0
\end{equation}
The last
term, $\:m^2/(1-x^2)$, must be 
removed in order to develop the solution as a polynomial in $\:x$.
This is achieved (as is well known) by writing the solution
$P(x)\:$ as the product of a known factor $\:(1-x^2)^\alpha\:$ and a
to-be-determined factor, $\:P^{\prime}(x)$, that is anticipated 
to be a polynomial in $\:x$:
\begin{equation}\label{prod}
P(x) = (1-x^2)^\alpha P^{\prime}(x)
\end{equation}

After substitution of the product form (\ref{prod}) and its derivatives, 
the differential equation for $\:P(x)\:$ 
(\ref{SEx}) becomes an equivalent equation for the 
dependent variable $\:P^{\prime}(x)$:
\begin{equation}\label{SEQ}
(1-x^2)\frac{{\rm d}^2P^{\prime}}{{\rm d}x^2} 
- (2 + 4\alpha) x \frac{{\rm d}P^{\prime}}{{\rm d}x} + 
\left(A - 2\alpha +
\left[\frac{4 x^2\alpha^2 - m^2}{1-x^2}\right]\right)P^{\prime} = 0
\end{equation}
The  denominator $\:(1-x^2)\:$ will cancel out if we
choose $\:\alpha\:$ as 
follows:
\begin{equation}\label{alpha}
4\alpha^2 = m^2 \quad\Rightarrow\quad \alpha 
= \pm \frac{\sqrt{m^2}}{2}\quad
\Rightarrow\quad \alpha = \pm \frac{|m|}{2}
\end{equation}
and we choose the positive square-root:
\begin{equation}\label{alphap}
\alpha = \mbox{} + \frac{|m|}{2}
\end{equation}
because choosing $\:\alpha = \mbox{} - |m|/2\:$ would make the solution
(\ref{prod})  infinite at the ends of the range: 
$x=\pm 1$.
 
With this choice for $\:\alpha$, the differential equation (\ref{SEQ})
for $\:P^{\prime}(x)$
becomes:
\begin{equation}\label{SEQF}
(1-x^2)\frac{{\rm d}^2P^{\prime}}{{\rm d}x^2} 
- 2(|m|+1) x \frac{{\rm d}P^{\prime}}{{\rm d}x} + 
\left[A - |m|(|m|+1)\right] P^{\prime} = 0
\end{equation}

We  proceed to substitute
a power series for $\:P^{\prime}(x)\:$ as follows:
\begin{equation}\label{powerQ}
P^{\prime}(x) = \sum_{i=0} a_i x^{i+k}
\end{equation}
and  equating the coefficient of {\em every power} of 
$\:x\:$ to zero (since the powers of $\:x$
are linearly independent of each other)  
produces a set of homogeneous linear equations
that determine the initial index  
$k$, the coefficients $\:a_i$, and the  eigenvalues of
the separation constant $\:A$.

Bypassing some details \cite{Heck}, it turns out that choosing 
$\:k$=$0\:$ produces all possible solutions, and
 thus we obtain the equation:
\begin{eqnarray}\label{ps5}
 \sum_{i=0} 
\left(\rule{0in}{0.2in}
a_i \left[A - (|m|+i)((|m|+i+1)\right]
+ a_{i+2} \left[(i+2)(i+1)\right]\right) x^i = 0 
\end{eqnarray}
Equating each power of $\:x\:$ to zero
produces a set of linear
equations that determine the coefficients of the power series 
$\{a_i:i=0,1,2\ldots\}\:$ which are
concisely written as the matrix equation:
\begin{equation}\label{mateqn}
\left(\begin{array}{ccccccc}
T_{00} & 0 & 2 & 0 & 0 & 0 & \cdots\\
0 & T_{11} & 0 & 6 & 0 & 0 & \cdots\\
0 & 0 & T_{22} & 0 &12 & 0 & \cdots\\
0 & 0 & 0 & T_{33} & 0 &20 & \cdots\\
0 & 0 & 0 & 0 & T_{44} & 0 & \cdots\\
0 & 0 & 0 & 0 & 0 & T_{55} & \cdots\\
\vdots & \vdots & \vdots & \vdots & \vdots & \vdots & \ddots
\end{array}\right)
\left(\begin{array}{c}a_0\\ a_1\\ a_2\\ a_3\\ a_4\\ a_5\\
\vdots\end{array}\right)
=
\left(\begin{array}{c}0\\ 0\\ 0\\ 0\\ 0\\ 0\\ \vdots\end{array}\right)
\end{equation}
which is summarized by the equation:
\begin{equation}\label{mateqnsum}
{\bf T.a} = {\bf 0}
\end{equation}
where the elements of the square matrix $\:{\bf T}\:$ are given by:
\begin{eqnarray}\nonumber
\rule{0in}{0.2in} T_{ii} = A-(|m|+i)(|m|+i+1)\quad\quad 
T_{ii+2} = (i+1)(i+2)\quad\\\label{matelem}
\rule{0in}{0.2in}
T_{ij} = 0 \ \ {\rm if\ } j\neq i {\rm\ or\ } j\neq i + 2\quad\quad\quad
\end{eqnarray}

\subsection{The Eigenvalues of $\:A$}

Since the matrix $\:{\bf T}\:$ is triangular, its determinant is equal 
to the
product of its diagonal elements, and hence the eigenvalues of $\:A$
(denoted by $\:A_i\:$ -- defined as being those values of $\:A\:$ that 
make the determinant of $\:{\bf T}\:$ zero) are obtained
by equating any one of the diagonal elements to zero: 
\begin{eqnarray}\label{eigens} 
& A_i = (|m|+i)(|m|+i+1)\ {\rm for}\ i=0,1,2,\ldots &
\end{eqnarray}
This expression for the eigenvalues of the separation constant $\:A\:$ 
shows that for a given value of $\:|m|\:$ 
(for the two values of $\:m$: $\:m=\pm|m|$) there is a set of values
of $\:A\:$ that increase quadratically with increasing values of $\:i$:
\begin{equation}\label{Aim}
A_i = (|m|+i)(|m|+i+1) = i^2 + i(2|m|+1) + |m|(|m|+1)\quad i=0,1,2\ldots
\end{equation}
We see that the smallest value of $\:A_i\:$ is $\:|m|(|m|+1)\:$ and that 
there is no
upper limit on $\:i\:$ nor upon the value of $\:A_i$.  The index $\:i\:$ 
is the 
(mathematically) natural
``quantum number'' for designating an eigenvalue $\:A_i$; the 
index $\:i\:$ 
is the degree of the polynomial in  $\:x$, $\:P^{\prime}(x)$, in
equation (\ref{powerQ}).

However, it has become customary  in physics to designate
an alternative quantum number, $\:\ell$, defined by:
\begin{eqnarray}\label{Al}
\ell = |m|+i\quad \Rightarrow \quad
A_i &=& (|m|+i)(|m|+i+1) = \ell(\ell+1)\\\nonumber
\ell &=& |m|,(|m|+1),(|m|+2)\ \ldots
\end{eqnarray}
Thus we see that  $\:\ell\:$  can take any
positive  value beginning at  $\:|m|\:$  for a given value of 
the quantum number  $\:m\:$  in  $\:\exp(im\phi)$.

Furthermore it has also become customary in physics, to reverse the
precedence of the relationship between $\:\ell\:$ and $\:|m|\:$ given in 
(\ref{Al}) by 
regarding $\:\ell\:$ as primary, with $\:m\:$ as the secondary 
quantum number:
\begin{equation}\label{lm}
\ell = 0,1,2\ldots\quad
A = \ell(\ell+1)\quad\quad 0\le |m|\le\ell  \quad\Rightarrow\quad
  -\ell\le m\le +\ell
\end{equation}
This physical viewpoint arises from atomic spectroscopy wherein
the energy of an atomic state depends upon $\:\ell$, but all 
$\:2\ell$+$1\:$ states differing only in $\:m\:$ are degenerate 
in the absence of an external magnetic field
\cite[p.4]{Tomonaga}.

For a given value of $\:\ell\:$ the minimum and maximum values 
of $\:m\:$ correspond
to $\:m_{min}=-\ell\:$ and $\:m_{max}=+\ell$.  These $\:2\ell$+$1\:$ 
values 
of $\:m\:$ 
correspond to polynomial degrees, $\:i$, differing by 1:
\begin{eqnarray}\nonumber
\begin{array}{|r|ccccccc|}\hline
i & \phantom{-}0 & \phantom{-}1 & \phantom{-}2 & \ldots & 2 & 
1 & 0\\\hline
m & -\ell &  -\ell+1 &  -\ell+2 & \ldots & \ell-2 & \ell-1 & \ell
\\\hline
\end{array}
\end{eqnarray}
and hence these $\:2\ell+1\:$ values of $\:m\:$ span an 
interval equal to $\:2\ell\:$  that is symmetrical about zero.  
This interval is necessarily a non-negative integer,
 $\:n\:$, and hence:
\begin{eqnarray}
2\ell = n\quad{\rm a\ non{\small\rm -}negative\ integer}
\end{eqnarray}
and hence the allowed values of $\:\ell\:$ are:
\begin{eqnarray}
\ell = \frac{n}{2}
\end{eqnarray}
where $\:n\:$ is any non-negative integer.  
When $\:n\:$ is even, $\:\ell\:$ is itself an integer, 
but when $\:n\:$ is odd,  $\:\ell\:$ is a half of an odd-integer.  
This is how both integer and
half-odd-integer values of $\:\ell\:$ and $\:m$ arise in the solution 
of the associated Legendre differential equation.  This argument is 
essentially identical with that used in the abstract theory of angular 
momentum \cite[pp.311]{Kaempffer} to deduce that both integer, 
and half-odd-integer, values of $\:\ell\:$ and $\:m\:$ are allowed.

\subsection{The Eigenfunctions $\:\{a_i:i=0,1,2\dots\}$}

Choosing  $\:T_{ii}=0\:$ in (\ref{matelem})
(i.e.~$A_i = [|m|+i][|m|+i+1]\:$) will lead to a solution in which 
the only non-zero coefficients are $\:a_i$, and {\bf all} {\em lower} 
coefficients
of the same parity (i.e.~odd or even).  The ratios of these non-zero
coefficients are given by the {\bf recursion relation}:
\begin{eqnarray}\label{rec2}& &
[A_i -(|m|+k)(|m|+k+1))]a_k + (k+1)(k+2) a_{k+2} = 0  \quad \Rightarrow 
\\\nonumber & &
a_k = - \frac{(k+1)(k+2)}{(i-k)(2|m|+i+k+1)} a_{k+2}\quad{\rm for\ } 
	k=(i-2),(i-4) \ldots \{1 {\ \rm or\ } 0\}
\end{eqnarray}
The recursion will terminate at\, $\:k=1$\, if\, $\:i$\, is odd, 
                         and at\, $\:k=0$\, if\, $\:i$\, is 
even.  Since $\:k<i\:$ the factors in the 
denominator of (\ref{rec2}) are always both positive, and hence 
consecutive terms of the power series
in\, $\:x$\, alternate in sign.  
This recursion relation (\ref{rec2}) is
written as beginning with the highest-index coefficient $\:a_i$, 
from which the lower-index coefficients are calculated.

\begin{table}[htb]
\caption{\bf Legendre Functions $\:P_{\ell}^{|m|}(x)\:$ 
($\ell=|m|+i$)}\label{legyfunc}
$$\begin{array}{|c|lllllll|}
\hline
|m| & \multicolumn{1}{c}{\rm factor\ } &
\multicolumn{1}{c}{0} &
\multicolumn{1}{c}{1} &
\multicolumn{1}{l}{i=2} &
\multicolumn{1}{l}{i=3} &
\multicolumn{1}{l}{i=4} &
\multicolumn{1}{l|}{i=5} \\\hline
\rule[-3mm]{0mm}{8mm}
0 & \phantom{(1.} 1 & 1 & x & 
1{\rm -}3x^2 & 
3x{\rm -}5x^3 & 
3{\rm -}30x^2{\rm +}35x^4 & 
15x{\rm -}70x^3{\rm +}63x^5 
\\
\rule[-3mm]{0mm}{8mm}
\frac{1}{2} & 
(1{\rm -}x^2)^{\frac{1}{4}}
 & 1 & x & 
1{\rm -}4x^2 & 
3x{\rm -}6x^3 & 
3{\rm -}36x^2{\rm +}48x^4 & 
15x{\rm -}80x^3{\rm +}80x^5
\\
\rule[-3mm]{0mm}{8mm}
1 & 
(1{\rm -}x^2)^{\frac{1}{2}}
 & 1 & x & 
1{\rm -}5x^2 & 
3x{\rm -}7x^3 & 
3{\rm -}42x^2{\rm +}63x^4 & 
15x{\rm -}90x^3{\rm +}99x^5
\\
\rule[-3mm]{0mm}{8mm}
\frac{3}{2} & 
(1{\rm -}x^2)^{\frac{3}{4}}
 & 1 & x & 
1{\rm -}6x^2 & 
3x{\rm -}8x^3 & 
3{\rm -}48x^2{\rm +}80x^4 & 
15x{\rm -}100x^3{\rm +}120x^5
\\
\rule[-3mm]{0mm}{8mm}
2 & (1{\rm -}x^2)       & 1 & x & 1{\rm -}7x^2 & 3x{\rm -}9x^3 
& 3{\rm -}54x^2{\rm +}99x^4 & 15x{\rm -}110x^3{\rm +}143x^5
\\
\rule[-3mm]{0mm}{8mm}
\frac{5}{2} & 
(1{\rm -}x^2)^{\frac{5}{4}}
 & 1 & x & 1{\rm -}8x^2 & 3x{\rm -}10x^3 
& 3{\rm -}60x^2{\rm +}120x^4 & 15x{\rm -}120x^3{\rm +}168x^5
\\
\rule[-3mm]{0mm}{8mm}
3 & (1{\rm -}x^2)^{\frac{3}{2}} & 1 & x & 1{\rm -}9x^2 & 3x{\rm -}11x^3
& 3{\rm -}66x^2{\rm +}143x^4& 15x{\rm -}130x^3{\rm +}195x^5
\\
\rule[-3mm]{0mm}{8mm}
\frac{7}{2} & 
(1{\rm -}x^2)^{\frac{7}{4}}
 & 1 & x & 1{\rm -}10x^2 & 3x{\rm -}12x^3 
& 3{\rm -}72x^2{\rm +}168x^4 & 15x{\rm -}140x^3{\rm +}224x^5
\\
\rule[-3mm]{0mm}{8mm}
4 & (1{\rm -}x^2)^{2} & 1 & x & 1{\rm -}11x^2 & 3x{\rm -}13x^3
& 3{\rm -}78x^2{\rm +}195x^4& 15x{\rm -}150x^3{\rm +}255x^5
\\
\rule[-3mm]{0mm}{8mm}
\frac{9}{2} & 
(1{\rm -}x^2)^{\frac{9}{4}}
 & 1 & x & 1{\rm -}12x^2 & 3x{\rm -}14x^3 
& 3{\rm -}84x^2{\rm +}224x^4 & 15x{\rm -}160x^3{\rm +}288x^5
\\
\rule[-3mm]{0mm}{8mm}
5 & (1{\rm -}x^2)^{\frac{5}{2}} & 1 & x & 1{\rm -}13x^2 & 3x{\rm -}15x^3
& 3{\rm -}90x^2{\rm +}255x^4& 15x{\rm -}170x^3{\rm +}323x^5
\\
\rule[-3mm]{0mm}{8mm}
\frac{11}{2} & 
(1{\rm -}x^2)^{\frac{11}{4}}
 & 1 & x & 1{\rm -}14x^2 & 3x{\rm -}16x^3 
& 3{\rm -}96x^2{\rm +}288x^4 & 15x{\rm -}180x^3{\rm +}360x^5
\\
\hline
\end{array}$$
\end{table}

Alternatively, the recursion can be written as beginning with 
$\:a_0\:$ or $\:a_1$:
\begin{eqnarray}\label{rec3}
 a_{k+2} = - \frac{(i-k)(2|m|+i+k+1)}{(k+1)(k+2)} a_k
\quad{\rm for\ } 
	k= 0{\ \rm or\ }1 \ldots (i-2)
\end{eqnarray}
This recursion relation was used to generate Table \ref{legyfunc}; the
coefficient of $\:x^0\:$ or $\:x^1\:$ was chosen to be positive and 
all of the coefficients were re-normalized to make them the smallest
set of integers for a given polynomial degree $\:i$.

Table \ref{legyfunc} displays a striking similarity between the 
well-known associated
Legendre functions for integer values of 
$\:\ell\:$ and $\:|m|$, 
and the 
newly discovered functions for half-odd-integer values.  
One 
difference between them, is that the factor 
$\:(1-x^2)^{\frac{|m|}{2}} = \sin^{|m|}\theta\:$ 
is an integer power of $\:\sin\theta\:$ when $\:m\:$ is an integer, 
but that it
has a factor of $\:\sqrt{\sin\theta}\:$ when $\:m\:$ is a 
half-odd-integer;
in the latter case the gradient, 
$\: {\rm d} P^{|m|}_{\ell}(\theta)/{\rm d}\theta\:$
is infinite at the limits: $\:\theta$=$0$, $\:\theta$=$\pi$, whereas in
the integer case these gradients are finite.
The {\em polynomial}, 
$\:P^{\prime |m|}_{\ell}(\cos\theta)$, 
involves
only integer powers of $\:x\,$=$\,\cos\theta\:$ in both cases.

\begin{table}[htb]
\caption{\bf Normalization Integrals $\:N^2_{\theta}\:$ of 
Legendre Functions 
$P_{\ell}^{|m|}(x)\:$ }\label{legfuncnorm}
$$\begin{array}{|c|ccccccc|}
\hline
|m| & 
i & 
0 &
1 &
2 &
3 &
4 &
5 \\\hline
\rule[-3mm]{0mm}{8mm}
0 & &
\frac{2}{1}  & \frac{2}{3} & 
\frac{8}{5}  & \frac{8}{7} & 
\frac{128}{9} & \frac{128}{11}
\\
\rule[-3mm]{0mm}{8mm}
\frac{1}{2} & &
\frac{\pi}{2} &
\frac{\pi}{8}  & \frac{\pi}{2} & 
\frac{9\pi}{32}  & \frac{9\pi}{2} & 
\frac{25\pi}{8} 
\\
\rule[-3mm]{0mm}{8mm}
1 & &
\frac{4}{3}   & \frac{4}{15} & 
\frac{32}{21} & \frac{32}{45} & 
\frac{768}{55} & \frac{768}{91}
\\
\rule[-3mm]{0mm}{8mm}
\frac{3}{2} & &
\frac{3\pi}{8} &
\frac{\pi}{16}  & \frac{15\pi}{32} & 
\frac{3\pi}{16}  & \frac{35\pi}{8} & 
\frac{75\pi}{32} 
\\
\rule[-3mm]{0mm}{8mm}
2 & &
 \frac{16}{15} & \frac{16}{105} & 
 \frac{64}{45} & 
 \frac{192}{385} & \frac{6144}{455}
  & \frac{2048}{315}
\\
\rule[-3mm]{0mm}{8mm}
\frac{5}{2} & &
\frac{5\pi}{16} &
\frac{5\pi}{128}  & \frac{7\pi}{16} & 
\frac{35\pi}{256}  & \frac{135\pi}{32} & 
\frac{945\pi}{512} 
\\
\rule[-3mm]{0mm}{8mm}
3 & &
\frac{32}{35} & \frac{32}{315} & 
\frac{512}{385}
& 
\frac{512}{1365}
&
\frac{4096}{315}
& 
\frac{20480}{3927}
\\
\rule[-3mm]{0mm}{8mm}
\frac{7}{2} & &
\frac{35\pi}{128} &
\frac{7\pi}{256}  & \frac{105\pi}{256} & 
\frac{27\pi}{256}  & \frac{2079\pi}{512} & 
\frac{385\pi}{256} 
\\
\rule[-3mm]{0mm}{8mm}
4 & &
\frac{256}{315} & \frac{256}{3465} & 
\frac{1024}{819}
& 
\frac{1024}{3465}
& 
\frac{16384}{1309}
& 
\frac{81920}{19019}
\\
\rule[-3mm]{0mm}{8mm}
\frac{9}{2} & &
\frac{63\pi}{256} &
\frac{21\pi}{1024}  & \frac{99\pi}{256} & 
\frac{693\pi}{8192}  & \frac{1001\pi}{256} & 
\frac{1287\pi}{1024} 
\\
\rule[-3mm]{0mm}{8mm}
5 & &
 \frac{512}{693} & \frac{512}{9009} &
\frac{4096}{3465}
  & 
\frac{4096}{17017}
&
\frac{32768}{2717}
& 
\frac{32768}{9009}
\\
\rule[-3mm]{0mm}{8mm}
\frac{11}{2} & &
\frac{231\pi}{1024} &
\frac{33\pi}{2048}  & \frac{3003\pi}{8192} & 
\frac{143\pi}{2048}  & \frac{3861\pi}{1024} & 
\frac{8775\pi}{8192} 
\\
\hline
\end{array}$$
\end{table}

\subsection{Normalization of the Spherical Harmonics}

The normalization constant, $\:N^2$, of $\:Y(\theta,\phi)\:$ 
(equation \ref{TFP}) 
is defined by:
\begin{equation}\label{normboth} 
N^2 = \int_0^{2\pi} \left|\Phi(\phi)\right|^2\, {\rm d}\phi\, 
\int_0^{\pi} \left[\Theta_{\ell}^{|m|}(\theta)\right]^2\, 
\sin\theta\,{\rm d}\theta
\end{equation}
The $\:\phi\:$ factor of $\:N^2\:$ is always $\:2\pi\:$ because 
$|\Phi(\phi)|^2$=$\exp(-{\rm i}m\phi)\exp({\rm i}m\phi)$=$1$.  However, since 
the domain of the functions for the half-odd-integer values of $m$, 
is $0$$\leq$$\phi$$<$$4\pi$, 
the integration over $\phi$ involved in (\ref{normboth}) should also,
in principle, be over this range $0$$\leq$$\phi$$<$$4\pi$.  This
 would only make the $\:\phi\:$ factor of $N^2$ equal to $\:4\pi\;$ rather than
$\:2\pi\:$, and so 
for the sake of consistency we choose
to have the same range of integration for both the integral and half-integral
values of $m$.

The $\:\theta\:$ factor of $\:N^2\:$ ($N^2_{\theta}$) is given by:
\begin{equation}\label{normlegy} 
N^2_{\theta} =
\int_0^{\pi} \left[\Theta_{\ell}^{|m|}(\theta)\right]^2\, 
\sin \theta\,{\rm d}\theta =
\int_{-1}^{+1} \left|P_{\ell}^{|m|}(x)\right|^2\, {\rm d} x
\end{equation}
One difference between the case of integer values of $\:\ell\:$, 
and that of half-odd-integer values, is that in the latter case 
$N^2_{\theta}\:$ has a factor of $\:\pi$, whereas in the integer case 
it is a rational fraction.  This is illustrated in 
Table \ref{legfuncnorm}, whose values correspond to normalizing 
the functions shown in Table \ref{legyfunc}.

\section{Discussion}

The fermion quasi-spherical Harmonics were discovered by Dr.~Ian Schlifer 
during a summer collaboration with Geoffrey Hunter several years ago 
\cite{Schlifer}.  More recent collaborative work involving the other 
authors of this article led to the recognition of the potential 
utility of these functions for modeling the magnetic field of 
fermions such as the electron \cite{Heck,Arslan}.

The simplicity of the functions and their similarity to the 
well-known spherical harmonics for integer $\:\ell\:$ and $\:|m|$
(Table \ref{legyfunc}), suggested that they might have been 
discovered many years ago.  However, a quite extensive literature 
search, including treatises on mathematical functions \cite{AandS} 
and on theoretical physics \cite{MandF}, did not reveal any 
previous presentation of them.

They have existed in principle as special cases of the 
hypergeometric function 
\cite[p.332 \& pp.561-562]{AandS}, but:
\begin{itemize}
\item the polynomial nature of their factors, 
$\:P^{\prime |m|}_{\ell}(\cos\theta)$, and 
\item their potential application as eigenfunctions 
of the spin \\
angular momentum of fermion particles,
\end{itemize}
have not previously been recognized.

\subsection{Mathematical Aspects}

It is beyond the scope of this article to attempt a comprehensive 
consideration
of these fermion spherical harmonics comparable with that in a standard
treatise \cite[pp.332-341]{AandS}.
Here we simply point
out some of the salient mathematical aspects to be considered in
adapting the general theory of Legendre functions 
to the fermion quasi-spherical harmonics:
\begin{itemize}
\item
The formul\ae\ which are apparently  undefined
when $\:\ell\:$ and $\:m\:$ are not integers (because they 
involve differentiation
with respect to $\:\theta$, $\:\ell\:$ or $\:m\:$ times), 
notably Rodrigues' Formula 
\cite[\S 8.6.18 p.334]{AandS}:
\begin{eqnarray}\nonumber
P^{0}_{\ell}(x) & = & \frac{1}{2^n n!}
\frac{{\rm d}^n(x^2-1)^n}{{\rm d}x^n}
\end{eqnarray}
and the  formula involving differentiation w.r.t. $m$
\cite[\S 8.6.6 p.334]{AandS}:
\begin{eqnarray}\nonumber
P^{|m|}_{\ell}(x) & = & (-1)^{|m|} (1-x^2)^{\frac{|m|}{2}}
\frac{{\rm d}^{|m|}P^{0}_{\ell}(x)}{{\rm d}x^{|m|}}
\end{eqnarray}
may  be applicable by the theory of 
semi-differentiation \cite[p.115 \& p.307]{Fcalc}.
\item 
The recurrence relation \cite[\S 8.5 p.334]{AandS}:
\begin{eqnarray}\nonumber
(\ell-|m|+1) P^{|m|}_{\ell+1} & = &
(2\ell+1) x P^{|m|}_{\ell} - (\ell+|m|) P^{|m|}_{\ell-1}
\end{eqnarray}
\end{itemize}
is  applicable to the case of  half-odd-integer values of
$\:\ell$\ and $\:|m|$; it is noteworthy that all the coefficients
in this recurrence relation: 
$(\ell-|m|+1)$, $(2\ell+1)$, $(\ell+|m|)$, 
are integers even when $\:\ell\:$
and $\: m\:$ are half-odd-integers.

This recurrence relation produces
a normalization and phase in which the {\em highest} power 
of $\:x\:$ in each
polynomial is {\em positive}, and in which the polynomial 
coefficients are
generally fractions.  A different normalization and phase was chosen
to construct Table \ref{legyfunc}
in order to display the regularity of the series of functions.
 
\subsection{Interpretation}
\label{interp}

The mathematical way of defining the half-odd-integer, 
quasi-spherical harmonics
to be proper (i.e.~single-valued) functions of the
angle  $\:\phi\:$, is to define its range to be 
$0\leq\phi\leq 4\pi$, since: 
\begin{eqnarray}\nonumber
\exp(i m [\phi+4\pi]) = + \exp(i m \phi)
\end{eqnarray}
when $\: m \:$ is half of an odd integer.
This is concordant with the well-known $\:4\pi\:$ symmetry 
of fermion 
wavefunctions; i.e.~the angle $\:\phi\:$ must transit
two complete circles for the wavefunction
to return  to its original 
value \cite[p.21 \& p.138]{Icke}, \cite[p.141]{Jones}.

However, with the range of $\:\phi\:$ redefined in this way, 
$\:\phi\:$ can no longer be regarded as one of the coordinate angles 
of the points on the surface of a sphere.  Rather it is an angular 
coordinate of the points on a double-sphere, for which the points 
on the outer surface of the sphere are different from the 
corresponding points on the inner surface \cite[p.419]{Kauffman}. 
The geometry
and topology of the double sphere 
can be modeled by the {\sl Dirac belt 
trick} \cite[p.417ff]{Kauffman}. 
Such a closed-on-itself surface with  
$\:4\pi\:$ periodicity
is the well-known M\"obius band \cite[pp.141-143]{Nash}, 
\cite[p.418]{Kauffman}.
Thus the fermion spherical harmonics are not true 
{\em spherical} harmonics,
but rather {\em quasi-spherical} harmonics.  

An alternative way of dealing with these functions is to define their domain
as the spherical range $0$$\leq$$\phi$$<$$2\pi$, but then to recognize
that they are double-valued functions of $\phi$.  This is the approach of
Bethe's theory of double-groups as discussed by Altmann 
\cite[chapter 13]{Altmann}.  It has also been based upon the theory 
of non-simply connected spaces \cite[\S 3.2 pp.35-38]{Morandi}.

In group-theoretical terms the integer $\:\ell\:$ and $\:m\:$ functions
are representations of $\:SO(3)\:$ while the half-integers functions are
representations of $\:SU(2)\:$ \cite[pp.35-41]{Morandi}, 
\cite[pp.123-130]{Mackey}, \cite[pp.140-142]{Jones}.

Regardless of whether one takes:
\begin{itemize}
\item the single-valued function on a 
non-spherical domain
approach, or
\item the double-valued function on a spherical domain approach,
\end{itemize}
the angle $\:\phi\:$ does not have a classical, physical interpretation.
Notwithstanding the long history of spin angular momentum 
\cite{Tomonaga},
its physical nature  remains something that is not easily
interpreted in terms of rotational motion in classical space-time.

\section*{Acknowledgements}

Professor Martin Muldoon of the York University Mathematics 
Department advised us about the conditions for the hypergeometric 
function to have a polynomial factor.

The  collaboration (in particular with Stoil Donev)
was made possible by a grant from the Natural Sciences and 
Engineering Research Council of Canada,
 by a stipend for Daniel Beamish from the Province of 
Ontario Summer Work-Study program, and by an assistantship 
to Paule Ecimovic from the York University Faculty of Graduate Studies.


\end{document}